\documentstyle[12pt]{article}
\begin{document}
\pagestyle{plain}
\newcommand{\be}{\begin{equation}}
\newcommand{\ee}{\end{equation}}
\newcommand{\bea}{\begin{eqnarray}}
\newcommand{\eea}{\end{eqnarray}}
\newcommand{\vp}{\varphi}
\newcommand{\pr}{\prime}
\newcommand{\sech} {{\rm sech}}
\newcommand{\cosech} {{\rm cosech}}
\newcommand{\psib} {\bar{\psi}}
\newcommand{\cosec} {{\rm cosec}}
\def\vs {\vskip .3 true cm}
\centerline {\bf Exact Ground State of several $N$-body}
\centerline {\bf Problems with an $N$-body Potential}
\vs
\centerline {\bf Avinash Khare$^*$}
\centerline {Institute of Physics, Sachivalaya Marg,}
\centerline {Bhubaneswar 751005, India.}
\vs
{\bf Abstract}

\vfill
electronic address : khare@beta.iopb.stpbh.soft.net
\eject
\section{Introduction}

Over the years, the exact solution of $N$-body problems has attracted 
considerable
attention because of its possible relevance in statistical mechanics as well as
in atomic, nuclear and gravitational many-body problems. Whereas exact solution
of several $N$-body problems in one dimension is known by now [1,2], a class of
exact solutions including the bosonic ground state have been obtained for several $N$-body
problems in higher dimensions when they are interacting via a 
harmonic oscillator
potential [3,4,5,6,7,8]. Clearly it is of considerable interest to 
discover other
exactly solvable $N$-body problems in one as well as in higher dimensions.

The purpose of this paper is to show that a class of exact solutions including
the bosonic ground state of all these $N$-body problems in higher dimensions 
can also be obtained in case they
are interacting via an $N$-body potential of the form
\be\label{1}
V({\bf r}_1, {\bf r}_2,..., {\bf r}_N) = -{e^2\over\sqrt{\sum_i {\bf r}^2_i}}
\ee
In this context, I may add that recently I have already obtained a class of
exact solutions of the $N$-anyon problem (in two dimensions) 
in case they are interacting 
via this $N$-body potential [9]. Further, last year, I also obtained the
complete bound state spectrum of the $N$-particle problem in one dimension 
in case they 
are interacting via a variant of the above potential as given by
[10,11,12]
\be\label{2}
V(x_1, x_2,..., x_N) = -{e^2\over \sqrt{\sum_{i,j}(x_i-x_j)^2}} \, .
\ee
Subsequently, Gurappa {\it et al.} [13] showed that the complete bound state 
spectrum can also be obtained in case an $N$-body potential of the form 
$\beta^2/{\sum_{i<j} (x_i - x_j)^2}$ is added either to the 
oscillator or to the $N$-body potential (2).

Based on all these results, I conjecture that whenever an $N$-body problem is solvable in
case these $N$ particles are interacting via an external one body ( or pairwise)
oscillator potential, the same $N$-body problem is also solvable in case they interact via the $N$-body potential as given by eq. (\ref{1}) or by 
\be\label{1a}
V({\bf r}_1, {\bf r}_2,...,{\bf r}_N) = -{e^2\over\sqrt{\sum_{i<j} ({\bf r}_i-
{\bf r}_j)^2}} \, .
\ee 
I further conjecture that in either case, one can also add an $N$-body potential of the form
\be\label{2a}
V({\bf r}_1, {\bf r}_2,..., {\bf r}_N) = -{\delta^2\over\sum_{i<j} 
({\bf r}_i -
{\bf r}_j)^2}
\ee
or its variant
\be\label{2b}
V({\bf r}_1, {\bf r}_2,..., {\bf r}_N) = -{\delta^2\over\sum_i {\bf r}^2_i} 
\ee
and the problem is still solvable except that the degeneracy in the spectrum is now much reduced. Clearly, it is necessary to examine 
the known solvable $N$-body 
problems with the oscillator potential and check if these conjectures are valid
or not.This is precisely what I propose to do in this paper.

The plan of the paper is as follows. In Sec.II, I discuss the $N$-body 
Calogero-Marchioro [3] model in $D$-dimensions [4] and show that {\it a la} 
the oscillator case, even for the $N$-body potential as given by eq. (\ref{1a}),
one can obtain some  
exact eigenstates including
the (bosonic) ground state. I also consider the Sutherland variant of the 
problem and as in the oscillator case, I obtain the exact (bosonic)
ground state when the $N$-bodies are interacting via the $N$-body potential as
given by eq. (1). In Sec.III, I discuss the two-dimensional model of Murthy 
{\it et al.} [5] in which they have obtained some exact eigenstates including the
(bosonic) ground state all of which show novel correlations. I show that 
similar exact solutions with novel correlations 
can also be obtained in case one replaces the one-body oscillator potential
by the potential as given in eq. (1). I also discuss the two-body problem in great
detail and show that as in [6], in this case too it is completely solvable. In
Sec.IV, I consider the $D$-dimensional generalization [7] of Murthy {\it et al's}
model [5,6] and show that exact solutions including the (bosonic) ground state can again
be obtained in case they are interacting via the $N$-body potential (1). In Sec.V,
I consider a Calogero type model in $D$-dimensions [8] which has only two-body
interactions and show that one can obtain some exact eigenstates including the
(bosonic) ground state in case the $N$-bodies are interacting via the $N$-body
potential (1). Finally, in Sec.VI, I obtain
the entire discrete bosonic spectrum in case the 
$N$-particles are interacting in $D$-dimensions purely 
via the $N$-body potential as given by (\ref{1a}). Further, in all the 
above cases
I show that the discrete spectrum can still be obtained 
even if one adds the $N$-body potential
as given by eq. (\ref{2b}) (or (\ref{2a})) 
to the 
oscillator or our $N$-body potential 
except that the degeneracy in the 
discrete spectrum is now much reduced.

\section{Calogero-Marchioro Model With $N$-body Potential}

Long time ago, in an effort to generalize the original Calogero model [2] to
dimensions higher than one, Calogero and Marchioro [3] considered a model in three
dimensions in which the N-particles are interacting via the two-body
and the three-body inverse square interactions as well as by the pairwise 
harmonic oscillator potential. They were able to obtain some exact eigenstates
including the bosonic ground state of the system. Recently, we [4] have been able
to generalize their work to arbitrary space dimensions. In particular we considered
the following $N$-body Hamiltonian in $D$-dimensions 
\be\label{3}
H = - {\hbar^2\over 2m}\sum^N_{i=1}{\bf \nabla}^2_i 
+ {g\hbar^2\over 2m} \sum^N_{i < j} {1\over {\bf r}^2_{ij}}
+{G\hbar^2 \over 2m}\sum_{i < j} {{\bf r}_{ki} \cdot {\bf r}_{kj}\over {\bf r}^2_{ki} 
{\bf r}^2_{kj}} +{m\omega^2\over 4}
\sum^N_{i<j} {\bf r}^2_{ij}
\ee
and showed that some eigenstates including the bosonic ground state for the 
system are 
given by 
\be\label{4}
\psi_{n_r} = (\prod\limits_{i < j}{\bf r}^2_{ij})^{\Lambda_D/2} exp( 
- {1\over 2}\sqrt{{1\over 2N}}\sum_{i<j}
{\bf r}^2_{ij}) L_{n_r}^{\Gamma_D}(\sqrt{{1\over 2N}}
\sum_{i <j} {\bf r}^2_{ij})
\ee
with energy
\be\label{5}
E_{n_r} = \sqrt{{N\over 2}} (2n_r +\Gamma_D+1)\, , \ n_r = 0,1,2,... \, .
\ee
Here ${\bf r}_i$ is the $D$-dimensional position vector of the $i$'th particle and
${\bf r}_{ij}={\bf r}_i-{\bf r}_j$ denotes the relative separation of the $i$'th
and $j$'th particles while $r_{ij}$ denotes its magnitude. 
In writing this exact solution, we have scaled all distances 
${\bf r}_i \rightarrow \sqrt{\hbar\over m\omega} {\bf r}_i$ and the energy is
measured in units of $\hbar\omega$. Throughout this paper, whenever we discuss
the exact solutions for the oscillator potential, we shall always be 
working with
these scaled coordinates and the energy will always be measured in units of 
$\hbar\omega$.
The parameters 
$g$ and $G$
characterize the (dimension-less) two-body and three-body coupling constants 
respectively while
$L_a^b$ denotes the Laguerre polynomial. Here $g > - 1/2$ to stop the {\it fall to
the origin}. Further, $\Lambda_D$ and $\Gamma_D$ are two parameters determined in
terms of the parameters of the Hamiltonian by
\be\label{6}
\Lambda_D\equiv \sqrt{G} = {1\over 2} 
\bigg [\sqrt{(D-2)^2+4g} - (D-2) \bigg ]
\ee
\be\label{7}
\Gamma_D = {1\over 2} \bigg [D(N-1)-2 + \Lambda_D N(N-1) \bigg ] \, .
\ee

Let us now consider the same many body problem as given by eq. (\ref{3}) but with
the oscillator potential being replaced by the $N$-body potential as given by 
eq. (\ref{1a}).
Throughout this paper, whenever we talk of the exact solutions with the $N$-body
potential as given by eqs. (\ref{1}) or (\ref{1a}), 
 we shall rescale all distances ${\bf r}_i \rightarrow 
{\hbar^2{\bf r}_i/ me^2}$ and measure energy in units of 
${me^4/ \hbar^2}$ so that $m, e, \hbar$ are all scaled away.

On substituting the ansatz
\be\label{9}
\psi = (\prod\limits_{i<j}{\bf r}^2_{ij})^{\Lambda_{D}/2} \phi(\rho)
\ee
in the Schr\"odinger equation for the potential (\ref{1a}) one obtains
\be\label{10}
\rho \phi''(\rho) +(2\Gamma_D+1)\phi'(\rho)
+({2\over\sqrt N}-
2\rho\mid E\mid ) \phi(\rho) = 0
\ee
where $\Gamma_D$ is as given by eq. (\ref{7}) while
\be\label{11}
\rho^2 = {1\over N}\sum_{i<j}{\bf r}^2_{ij} \, .
\ee
On further substituting 
\be\label{12}
\phi(\rho) = \ exp (- \sqrt{2\mid E\mid}\rho ) \chi(\rho)
\ee
it is easily shown that $\chi(\rho)$ satisfies the equation
\be\label{13}
\rho\chi''(\rho)+(2\Gamma_D+1-2\sqrt{2\mid E\mid} \rho) \chi'(\rho) +
[{2\over\sqrt{N}}-(2\Gamma_D+1)\sqrt{2\mid E\mid}] \chi(\rho)=0
\ee
whose solution is a Laguerre polynomial
\be\label{14}
\chi (\rho) = L_{n_r}^{2\Gamma_D} (2\sqrt{2\mid E\mid}\rho)
\ee
with the corresponding energy eigenvalues being 
(in units of ${me^4/ \hbar^2}$)
\be\label{15}
E_{n_r} = - {1\over 2N(n_r +\Gamma_D+{1\over 2})^2} \, .
\ee
Several comments are in order at this stage. 
\begin{enumerate}
\item  For $D = 1$ this expression agrees
with the one derived by us earlier [10]. 
\item For $n_r = 0$,
the eigenfunction 
\be\label{16}
\psi_0 =  (\prod\limits_{i < j}{\bf r}^2_{ij} )^{\Lambda_{D}/2} 
\ exp (-\sqrt{2\mid E_0\mid} \rho)
\ee
has no nodes besides those coming from the singular centrifugal potential which forces
the eigenfunction to vanish whenever the coordinates of any two particles
coincide. Thus $\psi_0$ corresponds to the ground state of the system with the
corresponding ground state energy being as given by eq. (\ref{15}) with 
$n_r = 0$. 
For $ n_r > 0$, we have radial excitations over the ground state.
\item As $g\rightarrow 0$, we see from eq. (\ref{6}) that also $G\rightarrow 0$
and the wave function (\ref{16}) becomes the ground state eigenfunction of
the hyperspherical ``Coulomb problem'' in $D$-dimensions without centrifugal barrier and
with Bose statistics. Thus the situation is different from the one dimensional 
problem [10] where
as $g\rightarrow 0$, the eigenfunction is the ground state of the ``Coulomb
problem'' but with Fermi statistics. We shall in fact see that in all the
higher dimensional many body problems $(D > 1)$, unlike the one dimensional
case, as the coupling is switched off, the eigenfunction corresponds to that of
Bose statistics. It is not clear whether this difference in statistics
between the one and the higher dimensions has any deeper physical significance.
\item The $N$-body problem is still solvable if apart from our $N$-body 
potential (\ref{1a}) we also add the potential (\ref{2a}) to the Hamiltonian 
(\ref{3}). In this case, on
substituting the ansatz (\ref{9}) in the 
Schr\"odinger equation for the combined potential yields 
\be\label{16a}
\phi''(\rho) +{(2\Gamma_D+1)\over\rho}\phi'(\rho)+({2\over\rho\sqrt N}
-{\delta^2\over N\rho^2} - 2\mid E\mid ) \phi (\rho) = 0 \, .
\ee
On following the steps as given by eqs. (\ref{11}) to (\ref{15}) it is easily
shown that the exact eigenstates are given by 
(in units of $me^4/{\hbar^2}$)
\be\label{16b}
E_{n_r} = - {1\over 2N(n_r +\gamma+{1\over 2})^2} 
\ee
\be\label{16c}
\phi(\rho) = {\rho}^{(\gamma - \Gamma_D)} exp (-\sqrt{2\mid E \mid} \rho)
L_{n_r}^{2\gamma} (2\sqrt{2\mid E \mid} \rho)
\ee
where
\be\label{16d}
\gamma = \sqrt{\Gamma_D^2 + {\delta^2/N}} \, .
\ee
\item In the same way, the $N$-body problem with the oscillator potential as
given by eq. (\ref{3}) is also solvable in case we also add the 
$N$-body potential 
(\ref{2a}) to it. In particular, it is easily shown that the corresponding 
exact eigenstates are
\be\label{16e}
E_{n_r} = {\sqrt {N\over 2}} [2n_r +1 + \gamma]
\ee
\be\label{16f}
\psi_{n_r} = (\prod\limits_{i < j}{\bf r}^2_{ij})^{\Lambda_D/2} exp( 
- {1\over 2}\sqrt{{1\over 2N}}\sum_{i<j}
{\bf r}^2_{ij}) {\rho}^{(\gamma -\Gamma_D)} 
L_{n_r}^{\gamma}(\sqrt{{1\over 2N}}
\sum_{i <j} {\bf r}^2_{ij}) \,.
\ee
\end{enumerate}

It may be added here that following Calogero and Marchioro [3], we can also
obtain a subset of the (spin-less) fermionic eigenfunctions in the special case
of $N$ = 3 and 4 when these particles are interacting via the $N$-body potential as 
given by eq. (\ref{1a}). In particular, it is easily shown that for $N = 3$, a
set of completely antisymmetric eigenstates in three space dimensions are 
\be\label{17}
\psi_{n_r} = ({\bf r}_{12}\times{\bf r}_{23})(\prod\limits_{i < j}{\bf r}^2_{ij})^{f/2} 
exp (-\sqrt{2\mid E\mid} \rho) L_{n_r}^{2F} (2\sqrt{2\mid E\mid} \rho)
\ee
\be\label{18}
E_{n_r} = - {1\over 6 (n_r +F+{1\over 2})^2}
\ee
where
\be\label{19}
f = \sqrt{G} = {3\over 2} \bigg [ \sqrt{1+{4g\over 9}} - 1 \bigg ], \  F = 3f+3 \, .
\ee
Generalization to $D$ dimensions is straight forward.

On the other hand for $N = 4$, a set of completely anti-symmetrical eigenstates
in three space dimensions are given by
\be\label {20}
\psi_{n_r} = [({\bf r}_{12}\times{\bf r}_{23})\cdot {\bf r}_{34} ] 
(\prod\limits_{i < j}{\bf r}^2_{ij})^{f/2} 
exp (-\sqrt{2\mid E\mid} \rho) L_{n_r}^{2F'} (2\sqrt{2\mid E\mid} \rho)
\ee
\be\label{21}
E_{n_r} = - {1\over 8 (n_r +F'+{1\over 2})^2}
\ee
where f is as given by eq. (\ref{19}) while $F' = 6f+5$.
Again, generalization to $D$ dimensions is straight forward.

I might add here that a la the oscillator case [4], we can also obtain the exact
ground state of the corresponding Sutherland variant of the problem. In 
particular, consider the Hamiltonian as given by eq. (\ref{3}) but with the
oscillator potential being replaced by the potential as given by eq. (\ref{1}).
It is easily shown that the exact ground state of the system is given by
\be\label{22}
\psi_0 = (\prod\limits_{i <j}{\bf r}^2_{ij})^{\Lambda_{D/2}} 
\ exp({-\sqrt{2\mid E \mid \sum_i
{\bf r}^2_i}})
\ee
where
\be\label{23}
E _0 = - {2\over (N-1)^2[1+N\Lambda_D]^2} \, .
\ee
One of the unsolved problem is whether one can map this problem (at least
in some specific dimension $D$) to some random matrix problem and obtain exact
results for the corresponding many-body theory. In this context it may be noted
that for $g$ = 2, the corresponding oscillator problem (see eq. (\ref{3})) 
in two dimensions
has been shown to be connected to the random matrix problem for the complex
matrices [4].

\section{Novel Correlations With an $N$-body Potential}

In a recent paper, Murthy {\it et al.} [5] have proposed a model in 
two dimensions
with two-body and three-body interactions. They were able to obtain the bosonic 
ground
state and a class of excited states of the system exactly by adding an external
(one body) harmonic oscillator potential. The model was constructed in such a 
way
that the solutions have a novel correlation of the from 
\be\label{24}
X_{ij} = x_iy_j - x_j y_i
\ee
built into them. Note that unlike the Jastrow-Laughlin form, $X_{ij}$
is a pseudo-scalar. However, unlike the Laughlin type of correlation,
this correlation is not translationally invariant unless the radial degrees of 
freedom are frozen. I now show that the bosonic ground state and the radial
excitations over it can also be obtained in case the oscillator potential is 
replaced by the $N$-body potential as given by eq. (\ref{1}). Further, I also
obtain the complete solution of the two-body problem. Notice that the two-body 
problem is quite nontrivial since the center of mass motion cannot be separated.

Following Murthy {\it et al.} [5], we start with the $N$-particle 
Hamiltonian (in 
the scaled variables) as given by 
\be\label{25}
2H = - \sum^N_{i=1}{\bf \nabla}^2_i
+g_1\sum^N_{i\not =j} {{\bf r}^2_j\over X^2_{ij}}
+ g_2\sum^N_{i\not =j\not =k} {{\bf r}_j \cdot {\bf r}_k\over X_{ij}X_{ik}}
-{2\over \sqrt {\sum {\bf r}^2_i}}
\ee
where $X_{ij}$ is as given by eq. (\ref{24}) while $g_1$ and $g_2$ are
dimension-less coupling constants of the two-body and the three-body
interactions respectively. 

It is easily shown that
\be\label{31}
\psi_0 (x_i,y_i) = \prod\limits^{N}_{i < j} \mid X_{ij} \mid^g 
exp (-\sqrt{2\mid E_0\mid \sum_i{\bf r}^2_i})
\ee
is the exact ground state of the system with the corresponding ground state
energy being (in units of $me^4/\hbar^2$)
\be\label{32}
E_0 = - {1\over 2 [gN(N-1)+N-{1\over 2}]^2}
\ee
provided $g_1$ and $g_2$ are related to $g$ by 
\be\label{28}
g_1 = g(g-1), \ g_2 = g^2 \, .
\ee
It may be noted that this
$\psi_0$ is regular for $g \geq 0 $ which implies that $g_1 \geq - 1/4, g_2 \geq 0$.

There is a neat way of proving that we have indeed obtained the ground state
by using the method of operators [14]. In particular, let us define the
operators
\bea\label{33}
Q_{x_i} & = & p_{x_i} + ig \sum^N_{j \not = i} {y_j\over X_{ij}} 
- {i \alpha x_i\over
\sqrt{\sum_j{\bf r}^2_j}} \nonumber \\
Q_{y_i} & = & p_{y_i} - ig \sum^N_{j \not = i} {x_j\over X_{ij}} 
- {i \alpha y_i\over
\sqrt{\sum_j{\bf r}^2_j}}
\eea
and their Hermitian conjugates $Q^+_{x_i}$ and $Q^+_{y_i}$. It is easy to see
that the Q's annihilate the ground state (\ref{31}) i.e.
$Q_{x_i}\psi_0 = Q_{y_i}\psi_0 = 0$. Further, the Hamiltonian (\ref{25})
can now be recasted in terms of these operators as
\be\label{34}
H = {1\over 2} \sum_i \bigg [Q^+_{x_i}Q_{x_i}+Q^+_{y_i}Q_{y_i} \bigg ]+ E_0
\ee
where $E_0 (= -{\alpha^2/2})$ is as given by eq. (\ref{32}). Since the operator on the 
right hand side is positive definite and annihilates the ground state
wave function (\ref{31}) hence $E_0$ must be the (bosonic) ground state
energy of the system. Note that as in the other two- ( and multi-) dimensional
problems, we are unable to obtain the fermionic ground state of the system
analytically. In fact, we do not know of any $N$-body problem in two or more
space dimensions whose fermionic
ground state has been analytically obtained. 

\subsection{A class of Excited States} 

We now show that {\it a la} the oscillator case, even in our case, a class of
excited states can be obtained analytically. To that end, we consider the
ansatz 
\be\label{35}
\psi (x_i,y_i) = \prod\limits^{N}_{i < j} \mid X_{ij}\mid^g 
exp(-\alpha\sqrt{\sum_i{\bf r}^2_i})
\phi (x_i,y_i)
\ee
On using eqs. (\ref{25}) and (\ref{35}) it is easily shown that $\alpha^2 =-2E$ 
while $\phi$ satisfies the eigenvalue equation
\be\label{36}
\bigg [ - {1\over 2}\sum_i{\bf \nabla}^2_i 
+\alpha \sum_{i=1} {{\bf r}_i \cdot {\bf \nabla}_i\over
\sqrt{\sum_j{\bf r}^2_j}}+{A\over\sqrt{\sum_j{\bf r}^2_j}} + g \sum_{i\not =j}
(x_j{\partial\over\partial y_i}-y_j{\partial\over\partial x_i}) \bigg ]\phi
= 0 
\ee
where
\be\label{37}
A = \alpha [gN(N-1)+N-{1\over 2}] -1
\ee
The exact solutions of this differential equation are best studied in terms
of the complex coordinates $z = x+iy, z^* = x-iy,$ and their partial derivatives
$\partial\equiv {\partial\over\partial z} = {1\over 2} ({\partial\over\partial x}-i{\partial
\over\partial y})$ and $\partial^* \equiv {\partial\over\partial z^*} = {1\over 2} ({\partial\over\partial x}+i{\partial
\over\partial y})$. In terms of these coordinates the differential equation
for $\phi$ take the form
\be\label{38}
\bigg [ - 2 \sum_i \partial_i\partial_i^* +\alpha\sum_i {z_i\partial_i+z_i^*\partial^*_i\over
\sqrt{\sum_j z_j z^*_j}} +{A\over \sqrt{\sum_j z_j z_j^*}}+ 2g \sum_{i\not = j}
{z_i\partial_i-z_i^*\partial^*_i\over
z_i z_j^*-z_jz^*_i} \bigg ]\phi = 0 \, .
\ee
It is worth remarking that $\phi $ is also an eigenstate of the total angular
momentum operator i.e. $L\Phi = l\Phi$. We now classify some exact solutions
according to their angular momentum.

(a) {\bf $l = 0$ Solutions}

Let us define an auxiliary parameter t by $ t = 2\alpha \sqrt{\sum_i z_iz_i^*}$ and
let $\phi = \phi (t)$. On using this in eq. (\ref{38}), the differential
equation for $\phi$ reduces to
\be\label{39}
t \phi''(t) +(b-t)\phi' (t) - a = 0
\ee
where $b = 2N(N-1)g+2N-1$ , and $a = A/\alpha$. The
allowed solutions are [15] confluent hypergeometric functions, $\phi (t) = M(a,b,t).$
Thus the polynomial solutions are obtained when $a = - n_r$ with $n_r$
being positive integer. Here, the subscript $r$ denotes radial excitations.
The corresponding eigenvalues are
\be\label{40}
E_{n_r} = - {1\over 2[n_r+gN(N-1)+N-{1\over 2}]^2} \, .
\ee
Note that all these states have zero angular momentum. 

(b) {\bf $ l > 0$ Solutions} 

Let $t_z = \sum_i z^2_i$ and let $\phi = \phi (t_z)$. All the mixed derivative
terms in eq. (\ref{38}) drop out and we obtain 
\be\label{41}
2\alpha t_z \phi' (t_z) + A\phi = 0
\ee
whose solution is $\phi (t_z) = t^m_z$ where $ m = - A/{2\alpha}$. Note that
$\phi(t_{z})$ is also an eigenfunction of the angular momentum operator with the
eigenvalue $l = 2m$ . Hence the energy eigenvalues in this case are given by
\be\label{42}
E_{l} = - {1\over 2[gN(N-1)+N+l-{1\over 2}]^2} \, .
\ee

(c) {\bf $l < 0$ Solutions}

Similarly let $t_z = \sum_i (z^*_i)^2$ and $\phi = \phi (t_{z^*})$. In this 
case $\phi$ satisfies the differential equation
\be\label{43}
2 \alpha t_{z^*} \phi' (t_{z^*}) + A\phi = 0
\ee
whose solution is $\phi(t_{z^*})=t^m_{z^*}$ where $m = - A/2\alpha$. In this
case $\phi(t_{z^*})$ is an eigenfunction of the angular momentum operator with the
eigenvalue $l = - 2m <0$. Thus the energy eigenvalues in this case are given by
\be\label{44}
E_{l} = - {1\over 2[gN(N-1)+N-l-{1\over 2}]^2} \, .
\ee

(d) {\bf Tower of Excited States}

We can now combine solutions in cases (b) and (c) (with nonzero $l$) with the solutions
in (a) and obtain an even more general class of excited states. For example, 
let us define
\be\label{45a}
\phi(z_i,z^*_i) = \phi_1(t) \phi_2 (t_z)
\ee
where $\phi_1$ is the solution with $l = 0$, while $\phi_2$ is the solution with
$l > 0$. The differential equation for $\phi_1$ is again a confluent hypergeometric 
equation and the energy eigenvalues are given by
\be\label{45}
E_{n_r,l} = - {1\over 2[n_r+gN(N-1)+N+l-{1\over 2}]^2}
\ee
One may repeat the same procedure to obtain exact solutions for a tower of 
excited states with $l < 0$. Combining all these states, it is then clear  
that the exact energy eigenvalues may be written in the form
\be\label{46}
E_{n_r,m} = - {1\over 2[n_r+gN(N-1)+N+2\mid m\mid-{1\over 2}]^2}
\ee
We thus see that for both $l > 0$ and $l < 0$, one has a tower of excited
states. Actually, the existence of a tower is a general result applicable
to all excited states of which the exact solutions shown above form a
subset. In particular, following the arguments given in Bhaduri {\it et al.} 
[6], let
us separate the coordinates $(x_i,y_i)$ into one radial coordinate
$t = 2\alpha \sqrt{\sum_i {\bf r}^2_i}$ as above and $2N-1$
angular coordinates collectively denoted by $\Omega_i$ say. Then 
eq. (\ref{36}) can be expressed as
\be\label{47}
t\phi''(t)+[2N(N-1)g+2N-1-t]\phi'(t)-[{A\over\alpha}
+{2\over\alpha t} T] \phi = 0
\ee
where $T = D_2 +gD_1$. Here 
$D_n$
is an $n$'th order differential operator which only acts on functions of 
the angles $\Omega_i$.
On further factorizing
\be\label{48}
\phi (x_i,y_i) = R(t) Y(\Omega_i)
\ee
where $Y$ is a generalized spherical harmonics on the (2N-1)-dimensional sphere $S^{2N-1}$
it is easily seen that $Y$ is an eigenfunction of the operator $T$. Let us
say that the corresponding eigenvalue is $\lambda$ (of course it must be 
admitted
here that the hard part of the problem is to find $\lambda$). On further
writing $R(t) = t^{\mu} \tilde {R}(t)$, it is easily shown that $\tilde {R}(t)$ satisfies
a confluent hypergeometric equation
\be\label{49}
t\tilde {R}''(t) +(b-t)\tilde {R}'(t) - a\tilde {R}(t) = 0
\ee
provided
\be\label{50}
\mu = \bigg ( [N(N-1)g+N-1]^2+4\lambda \bigg )^{1/2} - [N(N-1)g+N-1] \, .
\ee
Here $b = 2N(N-1)g+2N-1+2\mu$ while $a = \mu+A/\alpha$. We thus get normalizable
polynomial solutions of degree $n_r$ where $a = -n_r$. Here $n_r  = 0,1,2...$ denote the
number of radial nodes. Hence the energy eigenvalues are given by
\be\label{51}
E_{n_r} = - {1\over 2[n_r+gN(N-1)+N-{1\over 2}+\mu]^2} \, .
\ee
It may be noted that $\mu$ is in general a complicated function of the 
coupling constant $g$.
We thus see that for a given value of $\mu$ (which in general is unknown) 
there is an infinite tower of energy eigenvalues for which $(-E)^{-1/2}$
is separated by a spacing of one unit. This tower structure is a characteristic
of the $N$-body potential as given by eq. (\ref{1}) or eq. (\ref{3}) and will be
present in any $N$-body problem in $D$-dimensions. Note that a similar tower structure
also exists in the case of any $N$-body problem in $D$-dimensions if they are
interacting via an oscillator potential except that in that case, for a given value
of $\mu$, there is an infinite tower of energy eigenvalues separated by a
spacing of two units. {\it A la} Bhaduri {et al.} [6], in our case too, 
the tower structure
and the angular momentum are useful in organizing a numerical or analytical
study of the energy spectrum. In particular, note that, the radial quantum
number $n_r$, and the angular momentum $l$, are integers, and hence they cannot change as the
parameter $g$ is varied continuously.

Before ending this discussion it is worth pointing out that the class of
excited states can also be obtained if we add the $N$-body potential 
(\ref{2b}) to our $N$-body potential as given by eq. (\ref{1}). In 
particular, on proceeding from eq. (\ref{35}) and repeating the steps, 
instead of eq. (\ref{38}) one now has
\bea\label{51a}
\bigg [ - 2 \sum_i \partial_i\partial_i^* 
& + & \alpha\sum_i {z_i\partial_i+z_i^*\partial^*_i\over
\sqrt{\sum_j z_j z^*_j}} + 2g \sum_{i\not = j}
{z_i\partial_i-z_i^*\partial^*_i\over
z_i z_j^*-z_jz^*_i} \nonumber \\ 
& + & {A\over \sqrt{\sum_j z_j z_j^*}} +{\delta^2\over 2\sum_j z_j z_j^*}
\bigg ]\phi = 0 \, .
\eea
On following the steps as given above it is then easy to show that
the exact energy eigenvalues are 
now given by
\be\label{51b}
E_{n_r,m} = - {1\over 2[n_r+\gamma+{1\over 2}]^2}
\ee
where 
\be\label{51c}
\gamma = \bigg ({[gN(N-1)+N+2\mid m \mid -1]^2 + \delta^2} \bigg )^{1/2} \, .
\ee
The corresponding energy eigenstates are also easily written down.

It is worth pointing out that if instead we add the $N$-body potential 
(\ref{2b}) to the oscillator potential, then on following the steps as
given in [6], the exact energy eigenvalues 
can be shown to be
\be\label{51d}
E_{n_r,m} = 2n_r + 1 + \gamma \, .
\ee

\subsection{The Two-Body Problem: Complete Solution}

As in the oscillator case [6], we now explicitly show that the two-body problem is
integrable and exactly solvable when the two bodies are interacting via the
$N$-body potential as given by eq. (\ref{1}). As in [6] this is best done 
by going
over to the hyperspherical formalism in two dimensions [16,17].
It may be noted that the two-body problem is quite non-trivial here
since the center of mass cannot be separated.

The two body Hamiltonian is given by (see eq. (\ref{25}))  
\be\label{52}
H = - {1\over 2} ({\bf \nabla}^2_1+{\bf \nabla}^2_2) - {1\over\sqrt{{\bf r}^2_1+{\bf r}
^2_2}} + {g_1\over 2}{({\bf r}^2_1+{\bf r}^2_2)\over X}
\ee
where $ X = x_1y_2-x_2y_1$. Note that the three-body term is obviously not there in the
two-body problem ! 
The two body problem is best solved in the hyperspherical coordinates.
To that end, let us parameterize the coordinates ${\bf r}_1,{\bf r}_2$ in terms
of three angles and one length, $(R,\theta,\phi,\psi$) as follows:
\bea\label{53}
x_1+iy_1 & = & R (\cos\theta \cos\phi - i \sin\theta \sin\phi) exp (i\psi) \nonumber \\
x_2+iy_2 & = & R (\cos\theta \sin\phi + i \sin\theta \cos\phi) exp (i\psi) \, .
\eea
For a fixed $R$, these coordinates define a sphere in four dimensions with
radius $R(R^2=r^2_1+r^2_2)$ and $\theta, \phi$ and $\psi$ are in the interval
\be\label{54}
-\pi/4 \leq\theta \leq \pi/4 \, , \ -\pi/2 \leq\phi \leq \pi/2 \, , 
\ -\pi \leq \psi \leq \pi \, .
\ee
The important point to note is that $X = x_1y_2-x_2y_1 = 
{R^2\over 2} \sin(2\theta)$
depends only on $R$ and $\theta$ and is independent of $\phi$ and $\psi$. As a 
result the two integrals of motion are the angular momentum operator $L$
\be\label{55}
L = \sum_i (x_ip_{y_i}-y_ip_{x_i}) = - i {\partial\over\partial\psi}
\ee
and the supersymmetry operator
\be\label{56}
Q = i [x_2 {\partial\over\partial x_1}+y_2 {\partial\over\partial y_1}-x_1{\partial\over
\partial x_2}-y_1{\partial\over\partial y_2} ] 
= - i {\partial\over\partial\phi} \, .
\ee
Thus with SUSY, the two body problem as given by eq. (\ref{52}) is integrable with
the four constants of motion being the Hamiltonian $H$, the angular part of $H, L$ and
$Q$.

It is easy to check that the bosonic ground state has the quantum
numbers $l$ and $q$ of the angular momentum and SUSY operators equal to zero. In this
context it is worth noting that the eigenstates of the SUSY operator $Q$ are neither
symmetric nor asymmetric, unless the eigenvalue $q = 0$. After finding a simultaneous 
eigenstate of $H, L$ and $Q$, we can separate it into symmetric (bosonic) and 
asymmetric (fermionic) parts. Note that these symmetric and anti-symmetric 
parts are eigenstates of $Q^2$ but not of $Q$.

The two-body Hamiltonian in terms of the hyperspherical coordinates is given by
\be\label{57}
H = - {1\over 2} \bigg [{\partial^2\over\partial R^2}
+{3\over R} {\partial \over
\partial R}- {\Lambda^2\over R^2}+{2\over R} \bigg ] 
+ {2g_1\over R^2 \sin^2 (2\theta)}
\ee
where the operator $\Lambda^2$ is the Laplacian on the sphere $S^3$ and is given by
\be\label{58}
-\Lambda^2= {\partial\over\partial\theta^2} 
- {2 \sin(2\theta)\over \cos(2\theta)}
{\partial\over\partial\theta} 
+{1\over \cos^2 (2\theta)} [{\partial^2\over\partial\phi^2}+
2 \sin(2\theta) {\partial^2\over \partial\phi\partial\psi}
+{\partial^2\over\partial\psi^2}] \, .
\ee
As in the oscillator case, the interaction term i.e. 
$2g_1/{R^2 \sin^2(2\theta)}$ is
independent of $\phi$ and $\psi$. Hence the operators $L$ and $Q$ commute with the 
Hamiltonian (\ref{57}) (since they also commute with the noninteracting 
$(g_1=0)$ Hamiltonian). Hence, for all $g_1$, we label the states with the eigenvalues of
$L$ and $Q$. Each of these states is four-fold degenerate under parity $L\rightarrow -L, \ 
Q\rightarrow Q$ and the Hamiltonian is also invariant under parity. 
Hence the states labeled
by the quantum numbers ($l,q$) have the same energy as $(-l,q)$. 
The Hamiltonian is also invariant under the discrete transformation 
${\bf r}_1\rightarrow
-{\bf r}_1$ and  ${\bf r}_2\rightarrow {\bf r}_2$. Note that under this transformation
$L\rightarrow L$ and $Q\rightarrow -Q$. Thus the states labeled by the quantum
numbers ($l,q$) have the same energy as ($l,-q$). In this way we obtain the four-fold
degeneracy of the states. In fact we shall see below that the states with
($l,q$ ) have the same energy as ($q,l$) since interchanging $q$ and $l$ leaves the differential
equation invariant. As a result, one has in fact an eight-fold  degeneracy for the
levels for which $\mid q\mid$ and $\mid l\mid$ are non-zero and different from
each other. However, the degeneracy is only four-fold if $\mid l\mid = \mid q\mid
\not = 0$. Similarly, there is a four-fold degeneracy between the states $(\pm l,0)$
and $(0,\pm l)$ if $l\not = 0$. It must be noted here that the degeneracy that 
we are talking about is a subset of the degeneracy of the noninteracting
system.

Let us now try to solve the eigenvalue equation $H\psi = E\psi$ with $H$ being as
given by eq. (\ref{57}). It is easily seen that if we write
\be\label{59}
\psi (R,\theta,\phi,\psi) = F(R)\Phi (\theta,\phi,\phi)
\ee
then the eigenvalue equation separates into angular and radial equations.
In particular, the angular equation is given by  
\be\label{60}
[\Lambda^2 +{4g_1\over \sin^2 2\theta}]\Phi = \beta(\beta+2)\Phi
\ee
where $\beta\geq -1$, while the radial equation is given by
\be\label{61}
F'' (R) +{3\over R} F'(R)+(2E+{2\over R}-{\beta(\beta+2)\over R^2})F(R)=0 \, .
\ee
The radial equation is easily solved, yielding
\be\label{62}
F(R) = R^{\beta} exp (-\sqrt{2\mid E\mid} R) M(a,b,2\sqrt{2\mid E\mid}R)
\ee
where $b = 2\beta+3, a=3/2+\beta-1/{\sqrt{2\mid E\mid}}$ and
$M(a,b,x)$ is the confluent hypergeometric function. Demanding $a = -n_r$, where
$n_r$ is a positive integer, yields the bound state energy eigenvalues as
\be\label{63}
E = - {1\over 2(n_r+\beta+{3\over 2})^2} \, .
\ee
The tower structure of the eigenvalues built on the radial excitations of the ground
state is obvious from eq. (\ref{63}). 

It must be emphasized here that $\beta$ is still unknown and has to
be obtained by solving the angular equation (\ref{60}). We now note that the
angular equation is in fact identical in the oscillator and our case and it has 
been analyzed in great detail by Bhaduri {\it et al.} \cite{6}. We can therefore borrow their
results and draw conclusions about the value of $\beta$ and hence the spectrum as
given by eq. (\ref{63}). Some of the conclusions in our case are 
\begin{enumerate}
\item On substituting
\be\label{64}
\Phi(\theta,\phi,\psi) = P(x) exp(iq\phi) exp(il\psi)
\ee
where
\be\label{65}
P(x) = \mid x \mid^a (1-x)^b (1+x)^c \Theta^{a,b,c} (x)
\ee
with $x = \sin\theta, \ b = \mid l+q\mid/4, c = \mid l-q\mid/4, 
a(a-1)=g_1=g(g-1)$
and $l,q$ being integers, it is easily shown that $\Theta (x)$ satisfies 
the Heun's equation
\bea\label{66}
(1 & - & x^2) \Theta''(x) + 2[{a\over x}-(b-c)-(a+b+c+1)x]\Theta'(x) \nonumber \\
   & + & [ {(\beta+1)^2\over 4} - (a+b+c+{1\over 2})^2
+2{a(c-b)\over x}]\Theta(x)=0 
\eea
whose solutions are characterized by the so-called $P$-symbol \cite{bat}.
Note that unlike the hypergeometric case, Heun's equation has four regular singular
points. 
\item In case $b = c$ (i.e. either $l = 0$ or/and $q=0$), the Heun's eq. (\ref{66}) is
exactly solvable [6] with corresponding $\beta$ being
\be\label{67}
\beta = 2m+2a+4b \, , \ \ m = 0,1,2,... .
\ee
Thus in this case $E^{-1/2}$ varies linearly with $a$ (as is the case for the
exact solutions of the many-body problem). As pointed out in [5], this is an
example of a conditionally exactly solvable (CES) problem \cite{ddkv}.
\item There are quasi-exactly solvable (QES) \cite{ucks} polynomial 
solutions of degree $p (p \ge 1)$ in case
 there is a specific relationship between $a,b$ and $c$ . As a result, $\beta$
and hence $E^{-1/2}$ is nonlinear in $a$. In particular, in this case $\beta$
is given by 
\be\label{68}
\beta = 2a+2b+2c+2p \, .
\ee
It is worth emphasizing that the  equation is exactly solvable at an infinite
number of isolated points in the space of parameters $(a,b,c)$. These are 
isolated points because if we vary $a$ slightly away from any one of them, 
the equation is not exactly solvable since $b,c$ can only take discrete
values and hence cannot be varied continuously.
\item Bhaduri {\it et al.} [6] have done a detailed numerical, perturbative 
and large-$g$
analysis of $\beta$ as a function of $g$ both when $0\leq g < 1$ and also 
for large
$g$. Their analysis is also valid in our case. For example, their Figs. 1 and 2 are also valid
in our case except that in our case, Fig.1 represents a plot of $\beta+2n_r+2$ as a function
of $g$ while Fig.2 represents a plot of $\beta+2-2g$ vs. $g$. Most of their 
discussion goes through in our case with this obvious modification. 
\item The two body problem is also completely solvable in case we add 
the two-body
potential as given by eq. (\ref{2b}) to the Hamiltonian as given by 
eq. (\ref{52}). In this case, the entire discussion about the angular part
as given above is still valid. The only modification is in the radial 
equation as given by (\ref{61}). In particular, instead of (\ref{61}) we now
have
\be\label{68a}  
F'' (R) +{3\over R} F'(R)+(2E+{2\over R}
-{\beta(\beta+2)+\delta^2 \over R^2})F(R)=0 \, .
\ee
whose solution is
\be\label{68b}
F(R) = R^{\gamma} exp (-\sqrt{2\mid E\mid} R) M(a,b,2\sqrt{2\mid E\mid}R)
\ee
\be\label{68c}
E = - {1\over 2(n_r+\gamma+{3\over 2})^2} \, .
\ee
where $b = 2\gamma+3, a = \gamma +{3/2} - {1/{\sqrt{2\mid E \mid}}}$ with 
$\gamma = \sqrt{(1+\beta)^2 +\delta^2} - 1$, while $M(a, b, x)$ is the
confluent hypergeometric function.
\item If instead if we add the two-body potential (\ref{2b}) to the 
corresponding oscillator problem, then it is easily
seen that the angular part is again unchanged while the solution to the
radial part is given by 
\be\label{68d}
E_{n_r} = 2n_r+\gamma+1
\ee
\be\label{68e}
F(R) = R^{\gamma} exp (-R^2/2) M(a,b, R^2)
\ee
where $a = 1+(\gamma-E)/2, b = \gamma+2$.
\end{enumerate}

\section{Novel correlations in Arbitrary Dimensions}

In the last section we have obtained the exact bosonic ground state with Novel
correlation as well as radial excitations over it in case the $N$-particles 
in two dimensions also interact
through the $N$-body potential. The purpose of this section is to generalize these
arguments to arbitrary number ($D$) of dimensions. In this context it is worth
recalling that in case the $N$ particles are interacting via the oscillator 
potential, such a generalization has recently been carried out by Ghosh [7].
Following him, let us consider the following many-body Hamiltonian in $D$-dimensions
\bea\label{69}
H & = & -{1\over 2}\sum^N_{i=1}{\bf \nabla}^2_i
-{1\over\sqrt {\sum_{i=1} {\bf r}^2_i}}
+ {g_1\over 2}\sum_R{{\bf Q}^2_{i_2i_3...i_D}\over P^2_{i_1i_2...i_D}} \nonumber \\
  & + & {g_2\over 2} \sum_R {{\bf Q}_{i_2i_3...i_D}\times {\bf Q}_{j_2j_3...j_D}\over 
 P_{i_ii_2...i_D}P_{i_ij_2...j_D}}
\eea 
where $R$ denotes the sum over all the indices from 1 to $N$ with the restriction that no two indices can have the
same value simultaneously. Here 
\bea\label{70}
{\bf Q}_{i_2i_3...i_D} & = & {\bf r}_{i_2}\times{\bf r}_{i_3}\times...
\times {\bf r}_{i_D} \nonumber \\
      P_{i_1i_2...i_D} & = & {\bf r}_{i_1} \cdot {\bf Q}_{i_2i_3...i_D} \, .
\eea
Thus the Hamiltonian has $D$-body and $(2D-1)$-body interactions only. Note that both
${\bf Q}_{i_2i_3...i_D}$ and $P_{i_1i_2...i_D}$ are
antisymmetric under the exchange of particle coordinates. Further, 
$P_{i_1i_2...i_D}$ vanishes when the relative angle between any two of the
particles is zero or $\pi$. 

For simplicity, let us first discuss the $D=3$ case is some detail and then merely quote
the results for the general case of $D$-dimensions. It is easily seen that the exact bosonic
ground state in 3-dimensions is given by
\be\label{71}
\Psi_0 = \ \prod\limits_{R}  \mid P_{ijk}\mid^g 
exp ({-\alpha\sqrt{\sum_i{\bf r}^2_i}})
\ee
where $E_0 = - \alpha^2/2$ provided 
\be\label{71a} 
g_1 = {g\over 2} ({g\over 2} - 1) \, , \ g_2 = {g^2 \over 4} \, .
\ee
The corresponding ground state energy $E_0$ is given by 
\be\label{72}
E_0 = - {1\over {2[{3\over 2} gN (N-1) (N-2)+{3N\over 2}-{1\over 2}]^2}} \, .
\ee
The fact that this is indeed the ground state is easily shown as follows. 
Consider the following annihilation operators in three dimensions 
(note that ${\bf r}_i = x_i \hat {i}+y_i \hat {j}+z_i\hat {k})$
\bea\label{73}
A_{x_i} & = & p_{x_i} + {ig\over 2}\sum_S {({\bf Q}_{jk})_{x_i}\over P_{ijk}} -
{i\alpha x_i\over\sqrt{\sum_j{\bf r}^2_j}} \nonumber \\
A_{y_i} & = & p_{y_i} - {ig\over 2}\sum_S {({\bf Q}_{jk})_{y_i}\over P_{ijk}} -
{i\alpha y_i\over\sqrt{\sum_j{\bf r}^2_j}} \nonumber \\
A_{z_i} & = & p_{z_i} + {ig\over 2}\sum_S {({\bf Q}_{jk})_{z_i}\over P_{ijk}} -
{i\alpha z_i\over\sqrt{\sum_j {\bf r}^2_j}}
\eea
where $S$ denotes the sum over all the repeated indices from 1 to $N$ with the
constraint that any two indices cannot have the same value simultaneously.
It is easily shown that the Hamiltonian (\ref{69}) in 3-dimensions 
can be written down as
\be\label{74}
H = {1\over 2} \sum^N_{i=1}
\bigg (A^+_{x_i} A_{x_i}+A^+_{y_i}A_{y_i}+A^+_{z_i}A_{z_i} \bigg ) +E_0
\ee
Further, it is easily checked that the annihilation operators $A$'s annihilate the wave
function $\psi_0$ as given by eq. (\ref{71}) and hence $\psi_0$ is indeed the
ground state with the ground state energy being $E_0$ as given by 
eq. (\ref{72}).

The excited states can also be obtained in our case [7]. To that
end we write 
\be\label{75}
\psi (x_i,y_i,z_i) = \psi_0 (x_i,y_i,z_i)\phi(x_i,y_i,z_i)
\ee
where $\psi_0$ is as given by eq. (\ref{71}) except that $\alpha$ is now
given by $E = - \alpha^2/2$. On substituting this ansatz in the Schr\"odinger 
equation for the Hamiltonian (\ref{69}), 
it is easily shown $\phi$ satisfies the equation
\be\label{76}
\bigg [-{1\over 2}\sum_i{\bf \nabla}^2_i 
- {g\over 2} \sum {{\bf Q}_{jk} \cdot {\bf \nabla}_i\over
P_{ijk}}+\alpha\sum{{\bf r}_i \cdot {\bf \nabla}_i\over \sqrt{\sum{\bf r}^2_i}}
+{A\over\sqrt{
\sum{\bf r}^2_j}} \bigg ] \ \phi = 0
\ee
where
\be\label{77}
A = \alpha \bigg [ {g\over 2} N (N-1)(N-2)
+{3N\over 2}-{1\over 2} \bigg ] -1 \, .
\ee
On assuming that $\phi$ is a function of $t = 2\alpha\sqrt{\sum{\bf r}^2_i}$,
it is easily seen that eq. (\ref{76}) for $\phi (t)$ reduces to the 
confluent hypergeometric equation
\be\label{78}
t\phi''(t)+(b-t)\phi'(t) - a\phi(t)= 0
\ee
when $a = A/\alpha, \ b = {2(A+1)/ \alpha}$. Hence the admissible normalizable
solutions of eq. (\ref{78}) are $M(a = -n_r, b, t)$ 
with the corresponding energy eigenvalues being 
\be\label{79}
E_{n_r} = - {1\over 2[n_r+{g\over 2}N(N-1)(N-2)+{3N\over 2}-{1\over 2}]^2} \, .
\ee
Note again the tower structure and the fact that $(-E_{n_r})^{1/2}$ 
is linear in $g$ for the exact
solutions.

We can also obtain the exact solutions in case we add the $N$-body potential
(\ref{2b}) to the Hamiltonian (\ref{69}). In particular the energy eigenvalues
are now given by
\be\label{79a}
E_{n_r} = - {1\over 2[n_r+\gamma+{1\over 2}]^2} 
\ee
where 
\be\label{79d}
\gamma = \bigg ([{3N +gN(N-1)(N-2) \over 2} -1]^2 +\delta^2 \bigg )^{1/2} \, .
\ee
The corresponding eigenfunctions can be easily written down. 
if instead
one adds the $N$-body potential (\ref{2b}) to the oscillator potential, then
the energy eigenvalues are given by
\be\label{79b}
E_{n_r} = 2n_r +1 +\gamma \, .
\ee 
The corresponding eigenfunctions can again be easily written down.

The generalization of the above results to $D$-dimensions is straight forward. 
In particular, it is easily
shown that the exact bosonic ground state and the tower of excitations over it
in the case of the $D$-dimensional many-body Hamiltonian (\ref{69})
are given by 
\be\label{80}
\psi_n = \prod\limits_ R \mid P_{i_1i_2...i_D}\mid^g M(a=-n_r,b,t) 
exp({-\alpha\sqrt{\sum_i{\bf r^2_i}}})
\ee
where
\be\label{81}
E_{n_r} = - {\alpha^2\over 2} = - {1\over 2[n_r+{DN\over 2}
+g D (^N C_D)-{1\over 2}]^2}
\ee
provided
\be\label{82}
g_2 = ({g\over (D-1)!})^2, \ g_1 = {g\over (D-1)!} [ {g\over (D-1)!}-1]
\ee
so that
\be\label{83}
g = {(D-1)!\over 2} [ 1\pm \sqrt{1+4 g_1}]
\ee
It is interesting to note that whereas the relation between $g_1$ and $g_2$
and also between $g$ and $g_1$ is $D$-dependent, the allowed ranges of $g_1$ are 
independent of $D$ i.e. in any dimensions, whereas the solutions in the upper branch
are regular for $g_1\geq - {1\over 4},$ in the lower branch the regular solutions
are only allowed in the limited range $-{1\over 4} \leq g_1\leq 0$. Here
$a = A/\alpha, b = {2(A+1)\over \alpha}$ where 
\be\label{83d}
A = \alpha \bigg ( {D\over 2} [N+2g (^N C_D)]
-{1\over 2} \bigg )-1 \, .
\ee

The exact solutions are also obtained if we add the 
$N$-body potential (\ref{2b}) to the Hamiltonian (\ref{69}) and the 
exact energy eigenvalues are given by
\be\label{83a}
E_{n_r} = -{1\over 2[n_r +\gamma +{1\over 2}]^2}
\ee
where 
\be\label{83b}
\gamma = \bigg ([{ND\over 2} +gD (^N C_D) -1]^2 +\delta^2 \bigg )^{1/2} \, .
\ee
Similarly, if we add the $N$-body potential (\ref{2b}) to the Hamiltonian 
(\ref{69}) but with the oscillator potential
then the exact solutions are given by 
\be\label{83c}
E_{n_r} = 2n_r +1 + \gamma \, .
\ee
The corresponding eigenfunctions can be easily written down in both the cases.

\section{Calogero-Sutherland Type Models in Higher Dimensions with $N$ body 
Interaction}

In Sec.II, we have considered one possible generalization of the Calogero-
Sutherland type  models in higher dimensions. The key point there was to have
a long-ranged three-body interaction term. This is over and above the 
long ranged two-body interaction term which is present in one dimension.
Only then it was possible to obtain a class of exact solutions including the 
bosonic ground state. 
Another possible generalization of Calogero-Sutherland model to higher
dimensions was considered recently by Ghosh [8] when he introduced two 
models with
purely two-body long ranged interactions and in both the cases he was able to
obtain the exact bosonic ground state and radial excitations over it in case
the $N$-particle also interact via a (one-body) oscillator potential. In this
section we shall show that the exact ground state and radial excitations
over it can also be obtained in case the oscillator potential is replaced by
the $N$-body potential as given by eq. (\ref{1}).

Following Ghosh [8], let us consider the Hamiltonian
\be\label{84}
H = -{1\over 2} \sum^N_{k=1}{\bf \nabla}^2_k
-{1\over\sqrt{\sum_k {\bf r}^2_k}}+V_1(\beta)
+V_2(\beta)+W_3(\beta)
\ee
where
\bea\label{85}
V_1(\beta) & = & {\beta^2\over 2} g(g-1)\sum_{k\not = j}{\mid{\bf r}_k\mid^{2(\beta-1)}\over
(\mid{\bf r}_k\mid^{\beta}-({\bf r}_j\mid^{\beta})^2} \nonumber \\
V_2(\beta) & = & {g\beta \over 2}(D+\beta-2)\sum_{k\not = j}
{\mid{\bf r}_k\mid^{(\beta-2)}\over
(\mid{\bf r}_k\mid^{\beta}-({\bf r}_j\mid^{\beta})} \nonumber \\
W_3(\beta) & = & {\beta^2\over 2} g(g-1)\sum_{i\not = j\not = k}
{\mid{\bf r}_i\mid^{2(\beta-1)}\over
(\mid{\bf r}_i\mid^{\beta}-\mid{\bf r}_j\mid^{\beta})({\bf r}_i\mid^{\beta}-\mid {\bf r}_k\mid
^{\beta})} 
\eea
where $g$ is a dimension-less constant. Note that for $\beta = 1$ and 2 the three body
interaction term $W_3$ vanishes. Thus even though we give results for arbitrary
positive values of $\beta$, it must be noted that one has long ranged two-body
interaction alone only if $\beta$ = 1,2.

We first note that the Hamiltonian (\ref{84}) can be written as
\be\label{85a}
H = {1\over 2} \sum_i {\bf A}_i^+ \cdot {\bf A}_i + E_0
\ee
where the annihilation operators ${\bf A}_i$ are given by
\be\label{86}
{\bf A}_i = - i{\bf \nabla}_i + i\beta g \sum_{i\not = j}{\mid{\bf r}_i\mid^{\beta-2}\over (\mid
{\bf r}_i\mid^{\beta}-\mid {\bf r}_j\mid^{\beta}} {\bf r}_i 
- {{i\alpha \bf r}_i\over \sum_j
{\bf r}^2_j}
\ee
while the ground state energy $E_0 (= -{\alpha^2/2})$ is 
\be\label{87}
E_0 = -{1\over 2[{ND\over 2}+{g\beta\over 2} N(N-1)-1]^2} \, .
\ee
The fact that $E_0$ is the bosonic ground state energy is verified by noting
that each of the operators ${\bf A}_i$ annihilate the ground state wave
function
\be\label{88}
\psi_0 = \prod\limits_{i<j}(\mid{\bf r}_i\mid^{\beta}-\mid{\bf r}_j\mid^{\beta})^{g}
exp (-\sqrt{2\mid E_0\mid \sum_i r^2_i}) \, .
\ee

It is straight forward to obtain the radial excitation spectrum. In particular,
it is easily shown that the corresponding exact eigenfunctions are
\be\label{88a}
\psi = \prod\limits_{i<j}(\mid{\bf r}_i\mid^{\beta}-\mid{\bf r}_j\mid^{\beta})^{g}
M (a = -n_r, b, t) exp (-\sqrt{2\mid E\mid \sum_i r^2_i}) \, .
\ee
with the corresponding eigenvalues being
\be\label{87a}
E_{n_r} = -{1\over 2[n_r +{ND\over 2}+{g\beta\over 2} N(N-1)-1]^2} \, .
\ee
Here $t = 2\alpha\sqrt{\sum_i{\bf r}_i^2}, a = A/\alpha, b = 2(A+1)/\alpha$
while $E = -\alpha^2/2$ and $A = \alpha[ND/2 + g\beta N(N-1)/2 -1]$. 

One can also obtain the spectrum in case one adds the $N$-body 
potential (\ref{2b}) to the Hamiltonian (\ref{84}). In particular, it is 
easily shown that the corresponding energy eigenvalues are
\be\label{87b}
E_{n_r} = -{1\over 2[n_r+\lambda+{1\over 2}]^2}
\ee
where
\be\label{87c}
\lambda = \bigg ([{ND\over 2}+{g\beta N(N-1)\over 2}-1]^2 
+\delta^2 \bigg )^{1/2} \, .
\ee
The corresponding eigenfunctions can also be easily written down. Similarly,
in the oscillator case too, the spectrum can be written down in the presence
of the $N$-body potential (\ref{2b}).

\section{Complete Bosonic spectrum of an $N$-body problem in $D$ dimensions}

Over the years, the exact solution of an $N$-body problem in three 
(and even arbitrary)
space dimensions has attracted considerable attention because of its obvious implications
in several areas. So far, the only $N$-body problem which is completely
solvable in two and higher dimensions is when $N$-bosons are interacting pair-wise
via harmonic interaction. The purpose of this section is to show that there is
one more $N$-body problem in two and higher dimensions for which the complete
discrete spectrum can also be obtained. In particular I show that the full
bound state spectrum can also be obtained in case $N$-bosons are interacting via
the $N$-body potential as given by eq. (\ref{2a}).

In particular, consider the Hamiltonian
\be\label{89}
H = - {1\over 2} \sum^N_{i=1} {\bf \nabla}^2_i - 
{1\over\sqrt{\sum_{i<j}({\bf r}_i-{\bf r}_j)^2}} \, .
\ee
After the separation of the center of mass (which in this case moves as a free particle),
the relative problem is usually discussed in terms of the Jacobi coordinates ${\bf \zeta}_j (j=1,2,...,N-1)$
defined by
\be\label{90}
{\bf \zeta}_j = ({j\over j+1})^{1/2} [{1\over j} ({\bf  r}_i+...+{\bf r}_j)-{\bf r}_{j+1}] \, .
\ee
However, it is more convenient to consider the problem in hyperspherical coordinates $(\zeta,\omega)$
of the $(N-1)$-dimensional vector ${\bf \xi}$. Here the hyper-radius $\zeta (0\leq \zeta\leq \infty)$ is
given by
\be\label{91}
\zeta^2 = \sum^{N-1}_{j=1} \ {\bf \zeta}^2_j
\ee
while $\omega$ is a set of [$(N-1)D$-1] angular coordinates. With this choice, the
Hamiltonian takes the form
\be\label{92}
H = - {1\over 2} \bigg [ {\partial^2\over\partial\zeta^2} +{(N-1)D-1\over\zeta}{\partial\over
\partial\zeta} -{L^2(\omega)\over\zeta^2}\bigg ] - {1\over\sqrt N\zeta}
\ee
where $L^2(\omega)$ is the ``grand angular'' operator whose eigen functions
are the orthonormal hyper-spherical harmonics $Y_{[L]}(\omega)$ with eigenvalues
$-k[k+(N-1)D-2]$. Here $k$ = 0,1,2,... is the grand angular quantum number. It is
thus clear that if we write the eigenfunction $\psi$ in the form
\be\label{93}
\psi(\zeta,\omega) = \phi (\zeta) Y_{[L]}(\omega)
\ee
then in the $\zeta$ variable one has essentially a one dimensional Schr\"odinger
equation
\be\label{94}
\bigg [{\partial^2\over\partial\zeta^2}+{(N-1)D-1\over\zeta} 
{\partial\over\partial\zeta}
-{k[k+(N-1)D-2]\over\zeta^2} +{2\over {\sqrt N} \zeta} \bigg ] \phi (\zeta) 
= -2 E\phi(\zeta)
\ee
for the Kepler problem in $(N-1)D$-dimensions. The corresponding discrete energy
energy values and eigen functions are given by
\be\label {95}
E_{n_r,k} = - {1\over 2N[n_r+k+{(N-1)D-1\over 2}]^2}
\ee
\be\label{96}
\phi(\zeta) = y^k exp(-y/2) L^{2a}_{n_r} (y)
\ee
where $a = (N-1) D/2+k-1, y = 2 \sqrt{2\mid E\mid}\zeta$ and $n_r = 0,1,2,...$ is
the radial quantum number. 

It is worth pointing out that in eq. (\ref{92}), apart from the $N$-body
potential (\ref{1a}), the only other potential for which the Schr\"odinger
equation can be solved for all values of $k$ is the two-body harmonic
oscillator potential. Besides, in both the cases one can also add the $N$-body
potential (\ref{2a}) and the problem is still analytically solvable. In 
particular, instead of (\ref{95}) the spectrum is now given by
\be\label{96a}
E_{n_r,k} = -{1\over 2N[n_r+{1\over 2}+\lambda]^2}
\ee
where
\be\label{96b}
\lambda = \sqrt{a^2 + \delta^2} \, .
\ee 
The corresponding eigenfunctions can also be easily written down.
Note however that the degeneracy in the spectrum is now very much reduced.

Our $N$-body potential is in a way richer than the oscillator potential
in that here one not only has an infinite number of negative energy bound
states, but one also has positive energy scattering states. In particular, for $E > 0$,
the solution to the Schr\"odinger eq. (\ref{94}) is given by
\be\label{97}
\phi_i(\zeta) = exp ({ip\zeta}) \zeta^k 
F \bigg (k+{(N-1)D-1\over 2} - {i\over p}, (N-1)D+2K-1,-2ip\zeta \bigg )
\ee
where $E = p^2/2$. Hence the phase shifts for this problem are 
\be\label{99}
e^{2i\delta_p} = {\Gamma(k+{(N-1)D-1\over 2}-{i\over p})\over
\Gamma(k+{(N-1)D-1\over 2}+{i\over p})}
\ee
Unfortunately, we are unable to draw any simple conclusion about the $N$-particle
scattering from this expression of the phase shift. 

\section{Summary and Open Problems}

In this paper, we have discussed several $N$-body problems in two and higher
dimensions and have provided support to the 
conjecture that whenever an $N$-body problem is 
(partially) solvable
in case the $N$-bodies are interacting by the harmonic forces, then the same
problem will also be (partially) solvable in case they are 
interacting by an $N$-body
potential as given by eq. (\ref{1}) (or (\ref{1a})). 
Our conjecture is based on the following simple observation. In all the 
many-body problems with harmonic forces discussed in [2-8,13,17], after 
the short distance correlation etc. is taken out, the problem essentially
reduces to that of harmonic oscillator in dimensions higher than one. Now 
the only two problems for which the bound state spectrum can be analytically 
obtained for all partial waves in two and higher dimensions are the 
oscillator and the Coulomb problems. This strongly suggests that those $N$-body problems
for which the Schr\"odinger equation essentially reduces to Coulomb problem in 
two and higher dimensions, after the short distance correlation part etc. is 
taken out, must also be (partially) solvable. 
While we have no proof for
this conjecture, we have not come across any counter example either. 
We have also studied other models in one and two 
dimensions with this $N$-body potential [9,10] which also 
provide support to this conjecture.
It is clearly necessary to examine other solvable many-body problems with 
the harmonic forces and check if our conjecture is true in those cases or not.

Another well known fact in non-relativistic quantum mechanics is that both the
Coulomb and the oscillator problems in two and higher dimensions are also
exactly solvable for all partial waves if one adds a potential of the 
form $\hbar^2\delta^2/2m r^2$ to either one of them. Based on this observation, 
we have conjectured that for the $N$-body problems with the oscillator or our
$N$-body potential as given by eq. (\ref{1}) (or (\ref{1a})), 
one can also add the $N$-body potential as given by eq. (\ref{2b}) 
(or by (\ref{2a})) and the problem is still (partially) solvable but now the
degeneracy in the spectrum is much reduced. While we have no proof for this 
conjecture, we as well as Gurappa {\it et al} [13] have provided support to
this conjecture through several examples. Besides, we have not come across any
counter example either.

Apart from the oscillator and the $N$-body potential (\ref{1}), are there other 
potentials for which any of these $N$-body problems in higher dimensions 
can be (partially) solved? 
While we have no definite answer, it appears that all other potentials will
only be quasi-exactly solvable \cite{km} since we know of no other potentials
in non-relativistic quantum mechanics in two and higher dimensions for which
the bound state spectra can be analytically obtained for all partial waves. 

There are several common features in all the $N$-body problems that we have 
studied in this paper and before [9,12] as well as those studied with 
the oscillator 
potential. Some of these are as follows.
\begin{enumerate}
\item By now several $N$-body problems exist in two and higher dimensions for
which a class of eigenstates including the bosonic ground state 
have been exactly obtained in case they also interact either via
the oscillator or the $N$-body potential (\ref{1}). 
All these problems have long ranged two-body and
in most cases also the long ranged three-body interactions. In the special case
of one dimension, there are only long ranged two-body interactions and the 
complete bound state spectrum has been obtained in both the cases [2,10]. 
\item In all the $N$-body problems with the oscillator or the $N$-body potential
studied so far, one always obtains 
the tower of
states characteristic of the oscillator or the Coulomb problem as the case
may be.  
We conjecture that this will be a
general feature of any many-body problem with either of these forces. 
While we have no general proof, we have not come across any counter 
example either.
\item All the $N$-body problems in two and 
higher dimensions (that have been studied so far) with either the 
oscillator or the $N$-body potential 
are only partially 
solvable in the sense that while the bosonic ground state and radial 
excitations over it are known, the full bosonic spectrum is still unknown
(except the one studied in the last section or its oscillator 
analogue).
It is worth pointing out that the same is also
true in the case of $N$-anyons interacting either via the oscillator  
or via our $N$-body potential [17]. Further, 
for none of these problems, the ground 
state of $N$-fermions is analytically known. I conjecture that the same will 
be true for 
any other $N$-body problem in two and higher dimensions interacting via either 
of these potentials. 
\item If anyon
example is any guide, then it would seem that 
in the case of any $N$-body problem in two and higher dimensions with 
the oscillator potential, a part of the
spectrum  will be linear in the coupling constant (say $\Lambda_D$  in Sec.II) 
while a part of 
the spectrum will be nonlinear in the coupling constant. 
On the other hand, in the case of our $N$-body potential
(\ref{1}), a part of $(-E)^{-1/2}$ will be  
linear and a part will be nonlinear in the coupling constant.
So far, in all the examples discussed in the literature, one 
has only been able to obtain the linear
part of $E$ (or $(-E)^{1/2}$) in the case of the oscillator (or $N$-body) 
potential. While it is not clear if one has obtained the full linear
spectrum in all the cases or not, it is certainly true that so far 
exact solution has not been obtained for 
{\it even one nonlinear state} in any of the $N$-body problems in two 
and higher dimensions with either of the potentials. I believe that if 
one can obtain exact solution for even one nonlinear state in any of these 
problems, it will be a major breakthrough. 
\end{enumerate}
  
It must be added here though that the oscillator potential is perhaps more
useful than our $N$-body potential since whereas the clustering property holds
in the case of the oscillator potential it does not hold in the case of the
$N$-body potential. On the other hand, the $N$-body potential is in a way richer
than the oscillator potential in that unlike it, one has both
bound and continuous spectrum in this case. The most important and of course
difficult problem is to find some physical application of the $N$-body
potential. Hopefully, some application will be found in the near future.

\pagebreak

\end{document}